\documentclass{amsart}

\def\M{\mathcal{M}} \def\C{\mathbb{C}} \def\Z{\mathbb{Z}}
\def\N{\mathbb{N}}  \def\k{\kappa}
\def\calL{\mathcal{L}}  
\def\bibref[#1]{\cite{#1}} \newtheorem{theorem}{Theorem}[section]
\newtheorem{lemma}{Lemma}[section]
\newtheorem{corollary}{Corollary}[section]
\newtheorem{remark}{Remark}[section]
\newtheorem{definition}{Definition}[section]
\newtheorem{conjecture}{Conjecture}[section] \makeatletter
\def\@oddhead{\underline{\hbox to \textwidth{Solitons and
Almost-Intertwining Matrices\hfill Kasman/Gekhtman}}}
\let\@evenhead\@oddhead \def\@oddfoot{$\overline{\hbox to
\textwidth{\hfill \bf\thepage\hfill}}$} \let\@evenfoot\@oddfoot
\advance\footskip.5in \makeatother

\begin{document}


\title{Solitons and ``Almost Intertwining'' Matrices}

\author{Alex Kasman}
\address{Department of Mathematics\\College of Charleston}
\author{Michael Gekhtman}
\address{Department of Mathematics\\University of
Notre Dame}

\maketitle

\begin{abstract}
We define the algebraic variety of almost intertwining matrices to be
the set of triples $(X,Y,Z)$ of $n\times n$ matrices for which
$XZ=YX+T$ for a rank one matrix $T$.  A surprisingly simple formula is
given for tau-functions of the KP hierarchy in terms of such triples.
The tau-functions produced in this way include the soliton and
vanishing rational solutions.  The induced dynamics of the eigenvalues
of the matrix $X$ are considered, leading in special cases to the
Ruijsenaars-Schneider particle system.
\end{abstract}

\section{Introduction}

The KP hierarchy (cf.\ \bibref[Sato,SW]) is a well studied system of integrable
non-linear partial differential equations with Lax form
$$
\frac{\partial\mathcal{L}}{\partial
t_i}=\left[\mathcal{L},\mathcal{L}^i_+\right]\qquad i=1,2,3,\ldots
$$
for a monic, first order pseudo-differential operator $\mathcal{L}$.
In one of its formulations, the KP hierarchy is a set of bilinear
equations for the ``tau function'' $\tau(t_1,t_2,t_3,t_4,\ldots)$
depending upon infinitely many ``time variables'' $t_i$ ($i\in\Z_+$).
In this paper we will consider $\tau$-functions of the form:
\begin{equation}
\tau_M(t_1,t_2,\ldots):=\det(X e^{g(Z)}+e^{g(Y)})\label{eqn:tau}
\end{equation}
where $M=(X,Y,Z)$ is a triple of $n\times n$ constant complex matrices
and
the function $g$ is defined as
\begin{equation}
g(W):=\sum_{i=1}^{\infty} t_i W^i \qquad t_i\in\C.
\label{eqn:g}
\end{equation}
(To avoid issues of convergence, we will here consider only the case
in which all but a finite number of the parameters $t_i\in \C$ are
non-zero.)  

It is \textit{not} true that \eqref{eqn:tau} always gives the formula
for a function which satisfies the KP hierarchy.  For instance, as we
shall see from Remark~\ref{rem:2x2} and Theorem~\ref{thm:main}, in the
$2\times 2$ case formula \eqref{eqn:tau} is only a tau-function
if $\det[(XZ-YX)(Y-Z)]=0$.  On the other hand, among the
solutions one can obviously write this way are the \textit{$1$-soliton
solutions} which are the natural generalizations in this context of
the solitary wave from which the term ``soliton'' was coined by
Zabusky and Kruskal \bibref[ZK].  The \textit{standard}
$\tau$-function for the 1-soliton solution takes the form
\eqref{eqn:tau} where $M=(X,Y,Z)\in\C^3$ is any triple of scalar
constants.  (To exclude the degenerate cases we must further assume
that $X$ and $Y-Z$ are non-zero.)  This $\tau$-function describes a
single line soliton of the KP equation.  More generally, one may be
interested in $\tau$-functions of $n$-soliton solutions (``nonlinear
superpositions'' of $n$ different line solitons) or their rational
degenerations.  These $\tau$-functions are usually written in a form
that looks very different than \eqref{eqn:tau}.

Of course, when $X$, $Y$ and $Z$ are scalar as in the
1-soliton case, then the determinant which appears in \eqref{eqn:tau}
is unnecessary.  However, the main result of this paper is that the
$n$-soliton solutions also take the form \eqref{eqn:tau} and that they
arise in the case that $X$, $Y$ and $Z$ are three $n\times n$ matrices
satisfying the condition $\textup{rank}(XZ-YX)=1$.  In fact, this
same rank one condition provides not only the non-degenerate soliton solutions to the
KP hierarchy but also their rational degenerations.  Thus, we see that
the 1-soliton $\tau$-function is merely a special case of this much more
general formula.


\section{Almost Intertwining Matrices}

It is common to say that an operator $X$ intertwines the operators $Y$
and $Z$ if one has that 
\begin{equation}
XZ=YX.\label{eqn:realint}
\end{equation}

\begin{definition} Given three $n\times n$ matrices $X$, $Y$ and $Z$, we
define the \textit{rank $\k(X,Y,Z)$ to which $X$
intertwines $Y$ and $Z$} by the formula
$$
\k(X,Y,Z)=\textup{rank}(XZ-YX)=n-\dim \ker (XZ-YX).
$$
For fixed $k,n\in\N$ ($0\leq k\leq n$) define
$$
\M_n^k=\{(X,Y,Z)|\k(X,Y,Z)\leq k\}
$$
to be the set of all triples of $n\times n$ matrices $M=(X,Y,Z)$ such
that $\k(M)\leq k$.
\end{definition}

In most instances, one expects to find that $\k(X,Y,Z)=n$, its maximum
value.  For $\k(X,Y,Z)$ to be lower means that $X$ does, in fact,
intertwine $Y$ and $Z$ on the positive dimensional subspace
$\ker(XZ-YX)$.  In particular, when $\k(X,Y,Z)=0$, then $XZ=YX$ and so
$X$ does actually intertwine the other two matrices.
If $Y$ and $Z$ are not intertwined by $X$, then the best one could ask
for would be for $\k(X,Y,Z)$ to be equal to one, and so it seems
reasonable to say that they are \textit{almost} intertwined in this case.

\begin{remark} Note that a triple $(X,Y,Z)$ is in $\M_n^k$ precisely when the
$k\times k$ minor determinants of the matrix $XZ-YX$ all vanish.
Consequently, $\M_n^k$ has the geometric structure of an affine
algebraic variety in the $3n^2$-dimensional vector space of $n\times
n$ matrix triples.\end{remark}

The following elementary observations will be used to establish the connection
between almost intertwining matrices and solitons:
\begin{lemma}\label{lem:somesyms}{
\begin{itemize}
\item There is a natural $GL(n)\times GL(n)$ action on $\M_n^k$ given by
$$
(G,H)\in GL(n)\times GL(n):\qquad
(X,Y,Z)\in \M_n^k\mapsto (GXH^{-1},GYG^{-1},HZH^{-1})\in\M_n^k.
$$
which restricts on the diagonal to the natural $GL(n)$ action of
simultaneous conjugation
$$
G\in GL(n):\qquad (X,Y,Z)\in\M_n^k\mapsto
(GXG^{-1},GYG^{-1},GZG^{-1})\in\M_n^k.
$$

\item Let $\Lambda$ and $\Omega$ be $n\times
n$ matrices satisfying the commutation relationships
$$
[\Lambda,Y]=0\qquad [\Omega,Z]=0,
$$
then
$$
\k(X,Y,Z)=\k(\Lambda X\Omega,Y,Z).
$$
\end{itemize}}
\end{lemma}
\begin{proof}
Both claims are easily verified by noting that
$\k(X,Y,Z)\leq k$ if and only if
\begin{equation}
XZ=YX+\sum_{i=1}^k v_i\otimes w_i\label{eqn:tensor}
\end{equation}
for $n$-vectors $\{v_i\}$ and $\{w_i\}$ and that $\k(X,Y,Z)$ is
exactly the minimum $k$ for which such an equation exists.
\end{proof}

The main result of this section is the following lemma:

\begin{lemma}\label{lem:H}{Given three $n\times n$ matrices ${\hat X}$, $Y$ and $Z$,
let $H(a,b,c)\in\C[a,b,c]$ be the polynomial defined by
\begin{equation}
H(a,b,c)=H_1(a)H_2(b,c)-H_1(b)H_2(a,c)+H_1(c)H_2(a,b)\label{eqn:H}
\end{equation}
with
$$
H_1(a)=\det\left({\hat X}(aI-Z)+(aI-Y)\right)
$$
and
$$
H_2(a,b)=(a-b)\det\left({\hat X}(aI-Z)(bI-Z)+(aI-Y)(bI-Y)\right)
$$
If 
$\k({\hat X},Y,Z)\leq 1$ 
then $H(a,b,c)\equiv0$ is the zero polynomial.}
\end{lemma}
\begin{proof}
To say that $\k({\hat X},Y,Z)\leq 1$ is equivalent to saying that there exist
vectors $v$ and $w_1$ such that 
\begin{equation}
{\hat X}Z-Y{\hat X}=vw_1^{\top}.
\label{eqn:vw1}
\end{equation}
(In the case $\k({\hat X},Y,Z)=0$ one of these vectors is the zero vector.)
Also, merely for the sake of convenience, we introduce the notation
$$
Z_a=(aI-Z)\qquad \textup{and}\qquad Y_a=(aI-Y)
$$
and recall that $\textup{adj}(M)$ is the classical adjoint matrix
(i.e. $\textup{adj}(M)=\det(M)M^{-1}$ if $M$ is invertible).

Now, using \eqref{eqn:vw1} to eliminate ``${\hat X}Z$'', one can rewrite $H_1(a),
H_2(a,b)$ as $$H_1(a)=\det(Y_a ({\hat X}+I) + v w_1^{\top}),\ H_2(a,b)=(a-b)
\det(Y_a Y_b ({\hat X}+I) + Y_{a+b} v w_1^{\top} + v w_2^{\top}), $$ where $w_2^{\top}=w_1^{\top}
Z$.

Next, since $H(a,b,c)$ depends on ${\hat X}$ polynomially, it is enough
to prove, that $H(a,b,c)=0$ for almost all ${\hat X}$. Let us assume that
$\det({\hat X}+I)=\gamma\ne 0$. Then we can eliminate reference to ${\hat X}$ by writing
$$H_1(a)=\gamma \det(Y_a + v u_1^T),\ H_2(a,b)=\gamma (a-b) \det(Y_a Y_b  +
Y_{a+b} v u_1^{\top} + v u_2^{\top})$$
where
$$
u_1^{\top}=w_1\cdot ({\hat X}+I)^{-1}\qquad u_2^{\top}=w_2^{\top}\cdot
({\hat X}+I)^{-1}.
$$

Let us further re-write $H_2(a,b)$ as 
\begin{eqnarray*}
H_2(a,b) &=& (a-b)\gamma\det\left (Y_a Y_b  + Y_{a}
v u_1^{\top} + v u_1^{\top} Y_b +  v (u_2^{\top}+ u_1^{\top} Y)\right
)\\
 &=& (a-b)\gamma \det\left( (Y_a + v
u_1^{\top})(Y_b + v u_1^{\top}) + v u_2^{\top}\right )\ . 
\end{eqnarray*}
Finally, denote $Y-v u_1^{\top}$ by $M$. We obtain 
$$ H_1(a)=\gamma\det(M_a),
H_2(a,b)=\gamma(a-b)\det(M_a M_b + v u_2^{\top})\ .$$

Note that
$$
\det(M_a M_b +
v u_2^{\top})=\det(M_a)\det(M_b) + u_2^{\top}\mbox{adj}(M_a M_b)v=H_1(a) H_1(b)
(1 + u_2^{\top} M_a^{-1}M_b^{-1}v)  
$$ 
and since $[M_a,M_b]=0$ we also have that 
$$M_a^{-1}M_b^{-1}=
\frac{1}{a-b} (M_b^{-1} - M_a^{-1})\ .$$
Therefore, 
$$
\det(M_a M_b + v u_2^{\top}) = H_1(a) H_1(b) + \frac{
H_1(a) (u_2^{\top}\mbox{adj}(M_b)v) - H_1(b)
(u_2^{\top}\mbox{adj}(M_a)v)}{a-b}.
$$
So, using the notation $p(a)= a H(a) - u^{\top}\mbox{adj}(M_a)v$, we
see that $\det(M_aM_b+vu^{\top})$ is a \textit{Bezoutian} of the form
$$
\det(M_aM_b+vu^{\top})=
\frac{p(a) H_1(b)-p(b) H_1 (a)}{a-b}.
$$

Substituting $H_2(a,b)=\gamma (p(a) H_1(b)-p(b)H_1(a))$ into the
expression for $H(a,b,c)$ immediately yields $H\equiv0$.
\end{proof}

\begin{remark}\label{rem:2x2}The special case $n=2$ turns out to be surprisingly simple.
A quick calculation verifies that for \textit{arbitrary} $2\times2$ matrices
$\hat X$, $Y$ and $Z$ the polynomial $H(a,b,c)$ is given by the
formula
$$
H(a,b,c)=(a-b)(b-c)(c-a)\det[(\hat XZ-Y\hat X)(Y-Z)].
$$
\end{remark}

\section{Tau-Functions}

\subsection{Main Theorem}

It is easy to check that if $\k(M)=0$ then the formula for $\tau_M$
defined in \eqref{eqn:tau} \textit{is} a tau-function of the KP
hierarchy.  In fact, in this case in which \eqref{eqn:realint} is satisfied
one has
$$\tau_M(t_1,t_2,t_3,\ldots)=\det(X+I)\textup{exp}(\sum_{i=1}^{\infty} (\sum_{j=1}^n(\lambda_j^i) t_i)
$$
where
$\{\lambda_j\}$ are the eigenvalues of $Y$.
Since the function 
$$u(x,y,t)=2(\log \tau_M(x,y,t,0,0,\ldots))_{xx}=0$$
is the trivial solution to the KP equation, 
we say that $\tau_M$
is merely a gauge transformation of the \textit{trivial} tau-function.

Moreover, with $g$ defined as in \eqref{eqn:g} and $\tau_M$
defined by \eqref{eqn:tau}, we observe that this is \textit{still} a
$\tau$-function in the case $\k(M)=1$.  In fact, it is more
interesting in this ``almost intertwining'' case since we get
non-trivial soliton and rational solutions in this way.

\begin{theorem}\label{thm:main}
If $\k(M)\leq1$ for $M=(X,Y,Z)$ then the function
$$
\tau_M(t_1,t_2,\ldots)=\det(Xe^{g(Z)}+e^{g(Y)})\qquad
g(W)=\sum_{i=1}^{\infty} t_iW^i
$$
is a tau-function of the KP hierarchy with corresponding
(stationary) Baker-Akhiezer function
$$
\psi_M(x,z):=\frac{\det(X (zI-Z)e^{x Z}+ (zI-Y)e^{x Y})}{z^n\det(Xe^{x
Z}+e^{x Y})}e^{xz}.
$$
\end{theorem}
\begin{proof}
Given the semi-infinite vector 
$\vec t=(t_1,t_2,t_3,\ldots)$, we use the notation $\tau_M(\vec
t)=\tau_M(t_1,t_2,\ldots)$.  For an arbitrary constant $a$,
we define the semi-infinite vector $[a]=(a,a^2/2,a^3/3,\ldots)$.  Then, it is
sufficient to prove that the continuous function $\tau(\vec t)$ defined in
\eqref{eqn:tau} satisfies the Hirota equation in Miwa form (cf.\
\bibref[KWZ,Zabrodin])
\begin{eqnarray}
0 &=& (b-c)\tau(\vec t-[a^{-1}])\tau(\vec t-[b^{-1}]-[c^{-1}])\nonumber\\
&&-(a-c)\tau(\vec t-[b^{-1}])\tau(\vec
t-[a^{-1}]-[c^{-1}])\label{eqn:hirota}\\&&+
(a-b)\tau(\vec t-[c^{-1}])\tau(\vec t-[a^{-1}]-[b^{-1}])\nonumber
\end{eqnarray}
uniformly in $a$, $b$ and $c$ and for all $\vec t$.

However, from the definition we see that
\begin{eqnarray*}
\tau(\vec t-[a^{-1}]) &=&
\det(Xe^{g(Z)}e^{\ln(I-a^{-1}Z)}+e^{g(Y)}e^{\ln(I-a^{-1}Y)})\\
 &=& a^{-1}\det(e^{g(Y)})\det(\hat X(aI-Z)+(aI-Y))\\
 &=& a^{-1}\det(e^{g(Y)})H_1(a)
\end{eqnarray*}
where we have chosen $\hat X=e^{-g(Y)}Xe^{g(Z)}$ and used the notation
of Lemma~\ref{lem:H}.
Similarly,
$$
(a-b)\tau(\vec t-[a^{-1}]-[b^{-1}])=a^{-1}b^{-1}\det(e^{g(Y)})H_2(a,b).
$$
Consequently, \eqref{eqn:hirota} is equivalent to demonstrating that
the polynomial $H(a,b,c)$ in Lemma~\ref{lem:H} is zero in the case of
this $\hat X$, $Y$ and $Z$.  But, according to
the second result in Lemma~\ref{lem:somesyms} we have that $\k(\hat
X,Y,Z)=\k(X,Y,Z)\leq 1$ and so Lemma~\ref{lem:H} demonstrates that the
Hirota equation is satisfied.

Once we know that $\tau_M$ is a tau-function, the formula for $\psi_M$
is derived from simply using the ``famous Japanese formula'' \bibref[SW]:
$$
\psi_M(x,z)=\frac{\tau_M(x-z^{-1},-z^{-2}/2,-z^{-3}/3,\ldots)}{\tau_M(x,0,0,\ldots)}e^{xz}.
$$
Note that the numerator is simply $\tau_M(\vec t-[z^{-1}])$ with $\vec
t=(x,0,0,\ldots)$.  So, again expanding this in terms of the power
series for the logarithm we derive the desired expression for $\psi_M$.
\end{proof}

\begin{remark} Technically, although the function $\tau\equiv0$ solves the
bilinear equations of the KP hierarchy, it is not generally considered
to be a tau-function.  (In particular, there is no associated operator
$\mathcal{L}$ satisfying the Lax equation or function $u(x,y,t)$
satisfying the KP equation.)  In the preceding we have not been
careful to make certain that $\tau$ is non-zero.  In fact, one can
certainly choose $M\in\M_n^1$ so that $\tau_M=0$.  Consequently,
Theorem~\ref{thm:main} should be understood to say that \textit{if}
$\tau_M$ is non-zero (which is generally the case) \textit{then} it is
a KP tau-function.
\end{remark}

\begin{remark} Since the Baker-Akhiezer function $\psi_M$ in
Theorem~\ref{thm:main} has the property that $z^ne^{-xz}\psi_M$ is a
polynomial in $z$, it must be that $\tau_M$ is the tau-function of a
rank-one KP solution with a (singular) rational spectral curve.  In
particular, it must be a soliton solution or one of its rational
degenerations.  Well known consequences (cf.\
\bibref[cmbis,invprobs,W]) of this fact are the following:
\end{remark}
\begin{corollary}{Let $K=K_M(t_1,t_2,t_3,\ldots,\partial_{x})$ be the ordinary
differential operator determined by simply substituting the formal
symbol $\partial_x$ in for $z$ in the polynomial
$$
K(t_1,t_2,\ldots,z)=\frac{\det(X (zI-Z)e^{g(Z)}+ (zI-Y)e^{g(Y)})}{\det(Xe^{(
Z)}+e^{g(Y)})}.
$$
Then, equating $x$ and $t_1$,  $\calL_M=K\partial_x K^{-1}$ satisfies the Lax equations
$$
\frac{\partial}{\partial t_i}\calL=[\calL,(\calL^i)_+].
$$
Moreover, the function $u(x,y,t):=\frac{\partial^2}{\partial
x^2}\log\tau_M(x,y,t,0,0,\ldots)$ satisfies the KP equation
$$
\frac{3}{4}u_{yy}=\left(u_t-\frac{1}{4}(6uu_x+u_{xxx})\right)_x.
$$}
\end{corollary}

\begin{remark} It is well known and easily verified (cf.\ \bibref[SW]) that
multiplication by a function of the form $\exp({\sum \gamma_i t_i})$ takes
one tau-function to another having the same corresponding Lax operator
$\mathcal{L}$.  Such a change is often referred to as a ``gauge
transformation'' in KP theory.  Since $\det\exp g(Y)$ is a function of
this form with $\gamma_i=\sum \lambda_j^i$ (where $\lambda_j$ are the
eigenvalues of $Y$ counted according to multiplicity) it follows that:
\end{remark}
\begin{corollary}\label{cor:othertau}{For $M=(X,Y,Z)\in \M_n^1$,
$$
\hat\tau_M(t_1,t_2,\ldots)=\det(Xe^{g(Z)}e^{-g(Y)}+I)
$$
is also a KP tau-function differing from $\tau_M$ by only a gauge transformation.}
\end{corollary}

\begin{remark}{Since the tau-function and Baker-Akhiezer function are
defined as they are by determinants of $X$, $Y$ and $Z$,
simultaneously conjugating all three leaves the corresponding solution
unchanged.  Consequently, it would be possible to use
Lemma~\ref{lem:somesyms} to take the quotient of $\M_n^1$ by the
action of $GL$ and then would be natural to define
$\tau_{\overline{M}}$ for $\overline{M}\in\overline{\M_n^1}=\M_n^1/GL(n)$.
}
\end{remark}

\subsection{Special Cases}
\subsubsection{Gelfan'd-Dickii Hierarchies ($N$-KdV)}

The $N$-KdV or Gelfan'd-Dickii hierarchies are special classes of KP
solutions for which $\calL^N$ is an ordinary differential operator and hence is independent of the KP flows whose indices are multiples of $N$.  In
particular, we say a tau-function is an $N$-KdV tau-function if it
factors as $\tau=f\cdot g$ where
$$
\frac{\partial}{\partial t_{iN}}g=0\ \forall i\in\N, \qquad
\frac{\partial}{\partial t_1}f=0.
$$
In other words, except for a factor independent $t_1$, $\tau$ is
independent of $t_j$ for all $j$ that are multiples of $N$.

Let $\M_n^1(N)$ be the subset of $\M_n^1$
$$
\M_n^1(N)=\{(X,Y,Z)\in\M_n^1\ :\ Y^N=Z^N\}.
$$

\begin{theorem}{For $M\in\M_n^1(N)$, the corresponding tau-function
$\tau_M$ is a solution of the $N$-KdV hierarchy.
}\end{theorem}
\begin{proof}
If we consider only the dependence upon $t_1$ and $t_j$ ($j$ a multiple of $N$)
then
\begin{eqnarray*}
\tau_M&=&\det(Xe^{t_1 Z+t_j Z^j}+e^{t_1Y+t_j Y^j})\\
&=& \det(Xe^{t_1Z+t_j Z^j}+e^{t_1Y+t_j Z^j})\\
&=& \det(Xe^{t_1Z}+e^{t_1Y})\det(e^{t_j Z^j}).
\end{eqnarray*}
\end{proof}

For example, if we consider the restriction $Y=-Z$, then we are
looking for matrix pairs $(X,Z)$ satisfying
$$
\textup{rank}(XZ+ZX)=1.
$$
In this case, the formula \eqref{eqn:tau} will produce a
tau-function solution to the KdV hierarchy (independent of all even
time flows).  (cf.\ \bibref[Braden] where a special case of this is
presented in the context of integrable particle systems.)

\subsubsection{Solitons}\label{sec:solitons}

The $n$-soliton solutions to the KP hierarchy are identified by these
properties:
\begin{enumerate}
\item The BA function $\psi(x,z)$ when multiplied by a degree $n$ polynomial
$q(z)=z^n+\cdots$ has the form
$$
\overline{\psi}(x,z)=q(z)\psi(x,z)=(\sum_{i=1}^{n}a_i(x)z^i)e^{xz}.
$$
\item There are $n$ independent linear ``conditions'' satisfied
by $\overline{\psi}(x,z)$ of the form
$$
\alpha_i
\overline{\psi}(x,\lambda_i)+\beta_i \overline{\psi}(x,\mu_i)=0
\qquad 1\leq i \leq n
$$
(with $\lambda_i\not=\mu_i$).
\end{enumerate}

These solutions can be constructed from $\M_n^1$ by choosing the point
$M=(X,Y,Z)$ with 
$$
X_{ij}=\frac{\alpha_{i}}{\beta_j(\lambda_j-\mu_i)}
\qquad
Y_{ij}=\mu_i\delta_{ij}
\qquad
Z_{ij}=\lambda_i\delta_{ij}.
$$
This can be verified, for instance, by 
noting that because $[Y,Z]=0$, the tau-function $\hat\tau_M$ takes the
form (cf.~Corollary~\ref{cor:othertau})
$$
\hat\tau=\det(Xe^{g(Z)-g(Y)}+I).
$$
 For any index set
$J\subset \{1,\ldots, n\}$, the principal minor of $X
e^{g(Z)-g(Y)}$ can be written as 
$$
{\displaystyle
 \left (\prod_{i\in J}
\frac{\alpha_i}{\beta_i}\ e^{g(\lambda_i)-g(\mu_i)}\right) \det
\left (\frac{1}{\lambda_i - \mu_{i'}}\right)_{i,i'\in J}\ . }
$$
The latter determinant is a Cauchy determinant and is equal to
$$
{\displaystyle \prod_{i,i'\in J; i< i'}
\frac{(\lambda_i-\lambda_{i'})(\mu_{i}-\mu_{i'})}{(\lambda_i-\mu_{i'})(\mu_{i}-\lambda_{i'})}
\prod_{i\in J}\frac{1}{\lambda_i-\mu_{i}} }
$$
Setting $c_i=\frac{\alpha_i}{\beta_i (\lambda_i-\mu_{i})}$, we
obtain
$$
{\displaystyle \hat\tau= \sum_{J\subset \{1,\ldots, n\}} \prod_{i\in
J} c_i \ e^{g(\lambda_i)-g(\mu_i)} \prod_{i,i'\in J; i< i'}
\frac{(\lambda_i-\lambda_{i'})(\mu_{i}-\mu_{i'})}{(\lambda_i-\mu_{i'})(\mu_{i}-\lambda_{i'})}
}
$$
which coincides with the known formula for this $n$-soliton solution
of the KP hierarchy (cf.~\bibref[Kac]).

\subsubsection{Polynomial $\tau$-functions and rational solutions}

Clearly, in the case that $Y$ and $Z$ are chosen to be nilpotent, the
definition of $\tau_M$ produces a \textit{polynomial} in the time
variables $t_i$.  It is perhaps of greater interest to note that one may also get tau-functions
that are -- up to a gauge transformation -- polynomial in $t_1$ but an
infinite series if all $t_i$ are considered.

For example, choosing
$$
X=\left(\begin{matrix}1&1&0\\
1&0&0\\
1&0&0\end{matrix}\right)
\qquad
Y=\left(\begin{matrix}\lambda&1&0\\
0&\lambda&1\\
0&0&\lambda\end{matrix}\right)
\qquad
Z=\left(\begin{matrix}\lambda&0&0\\
1&\lambda&0\\
0&1&\lambda\end{matrix}\right)
$$
leads to (after a gauge transformation to remove an exponential factor):
$$
\tau(x,y,t,0,0,\ldots)=1+(-3\lambda+3\lambda^2)t+\frac{9}{2}\lambda^4t^2+\frac{x^2}{2}+(6\lambda^3t+2\lambda-1)y+2\lambda^2y+(1+3\lambda^2t+2\lambda
y)x.
$$

Such solutions are well known and have been studied
in previous papers (cf. \bibref[AMM,cmbis,Kr,W,Z]).  However, one
should especially compare the present approach with that in \bibref[W2]
where these ``vanishing rational KP solutions'' are produced from matrix pairs
$(X,Z)$ satisfying $\textup{rank}(XZ-ZX+I)=1$.  The main results in
that paper concern the induced dynamics of the eigenvalues which
behave as particles in a Calogero-Moser particle system.  So, it may
be of interest to similarly investigate the dynamics of the
eigenvalues associated to almost intertwining matrices.

\section{Eigenvalue Dynamics}

One of the most interesting things about the Ruijsenaars-Schneider particle system \bibref[Braden,RS,R] is its connection to soliton tau-functions.  Specifically, certain KP tau-functions can be written as
$$
\tau(t_1,t_2,\ldots)=\det(X+I)
$$
where $X=X(t_1,t_2,t_3,\ldots)$ is a matrix whose eigenvalues move according to the Ruijsenaars-Schneider Hamiltonian.  

In this section we similarly study the dynamics of eigenvalues of time dependent matrices in the context of almost intertwining matrices to both reproduce and extend known results about the RS system and its connection to solitons.

\subsection{Solitons and a Matrix Flow}

\begin{theorem}{The vector fields $V_i$ on the space of $n\times n$
matrix triples defined by
\begin{equation}
V_i(X_0,Y,Z)=(X_0Z^i-Y^iX_0,0,0).
\end{equation}
are tangent to the manifold $\M_n^1$ and induce the flows in the
variables $t_i$ parametrized as
\begin{equation}
M_t=(X_t,Y,Z)=(e^{-g(Y)}X_0e^{g(Z)},Y,Z).\label{eqn:X_t}
\end{equation}}
\end{theorem}
\begin{proof}
Note that the flows specified have the stated vector fields and that
$$
X_tZ-YX_t=e^{-g(Y)}(X_0Z-YX_0)e^{g(Z)}
$$
is a rank one matrix if $X_0Z-YX_0$ is.
\end{proof}

\begin{remark} Given a parametrized flow $(X_t,Y,Z)\in\M_n^1$ as above, the
function $\hat\tau_M=(X_t+I)$ is another way to write the gauge
transformed tau-function from Corollary~\ref{cor:othertau} with $M=(X_0,Y,Z)$.
\end{remark}


\subsection{General Equations for Eigenvalue Dynamics}

Given any matrices $X_0,Y,Z$ such that $\k(X_0,Y,Z)=1$ let us define $X=X_t$ according to \eqref{eqn:X_t}.  If we denote the eigenvalues of $X_t$ by $\{Q_i(t)\}$ ($1\leq i \leq n$), to what extent can we describe their dynamics by intrinsic equations (depending only on $Q_i$ and their derivatives)?

In what follows we will only be considering the flow under the first time parameter $t_1$, but will write simply $t$ in order to simplify exposition and will use a "dot" to indicate differentiation with respect to this parameter.

We define vectors $v$ and $w$ by the formula
\begin{equation}
(X_0 Z- Y X_0) = vw^{\top}\label{eqn:vw}
\end{equation}
and so we have the equations of motion
\begin{equation}
\dot X=vw^{\top}\qquad \dot Y=0 \qquad \dot Z=0.\label{eqn:motion1}
\end{equation}

For convenience we introduce the (time dependent) matrix $U$ which diagonalizes $X$ and the logarithms of the eigenvalues $q_i$
\begin{equation}
Q=UXU^{-1}=\left(\begin{matrix}Q_1 & 0 &0 & \cdots\\
0 & Q_2 & 0 & 0 & \cdots \\
0&0& Q_3 & 0 & \cdots\\
\vdots &&&\ddots \end{matrix}\right) \qquad q_i=\ln(Q_i)
\label{eqn:Q}
\end{equation}
and define in analogy to \eqref{eqn:vw} the matrices and vectors
\begin{equation}
\hat Y=UYU^{-1}\qquad \hat Z=UZU^{-1} \qquad \hat v=Uv\qquad\hat w=wU^{-1} \label{eqn:hats}
\end{equation}
so that 
\begin{equation}
Q \hat  Z-\hat YQ=\hat v \hat w^{\top}.
\end{equation}
Note, in particular, that looking at an individual element of this last equation yields
\begin{equation}
Q_i \hat Z_{ij}-Q_j \hat Y_{ij}=\hat v_j\hat w_i.\label{eqn:linear1}
\end{equation}

Now, defining $M=\dot UU^{-1}$ we have in analogy to \eqref{eqn:motion1}
\begin{equation}
\dot Q=[M,Q]+\hat v \hat w^{\top}\qquad \dot{\hat Y}=[M,\hat Y]\qquad
\dot{\hat Z}=[M,\hat Z].
\label{eqn:motion2}
\end{equation}

Since $Q$ and $\dot Q$ have no off diagonal elements, we get from \eqref{eqn:motion2} that
\begin{equation}
\dot Q_i=\hat v_i \hat w_i=\dot q_ie^{q_i}
\label{eqn:dotqs}
\end{equation}
and
\begin{equation}
M_{ij}=\frac{\hat v_j\hat w_i}{Q_i-Q_j}\qquad (i\not=j).
\label{eqn:Mij}
\end{equation}

It turns out to be especially useful to write the equations of motion in terms of $q_i$ rather than $Q_i$ because then we find by multiplying \eqref{eqn:motion2} by $Q^{-1}$ that
\begin{equation}
\left(\begin{matrix}
\dot q_1 & 0 & \cdots \\
\vdots & \ddots &\\
0 & \cdots &\dot q_n\end{matrix}\right)
=\dot QQ^{-1}=M-QMQ^{-1}+Q\hat ZQ^{-1}-\hat Y.
\label{eqn:qdotversion}
\end{equation}
Since $Q$ is diagonal, $M-QMQ^{-1}$ has no diagonal and $Q\hat ZQ^{-1
}-\hat Y$ has the same diagonal as $\hat Z-\hat Y$ and so
\begin{equation}
\dot q_i=\left(\hat Z-\hat Y\right)_{ii}
\label{eqn:motion3}
\end{equation}

Finally, we can differentiate \eqref{eqn:motion3} and use \eqref{eqn:linear1}, \eqref{eqn:dotqs}\ and \eqref{eqn:Mij} to find the equation of motion
\begin{eqnarray}
\ddot q_i&=&([M,\hat Z-\hat Y])_{ii}
\\
&=&\sum_{k\not=i}\left(M_{ik}(\hat Z_{ki}-\hat Y_{ki})-M_{ki}(\hat Z_{ik}-\hat Y_{ik})\right)\\
&=&\sum_{k\not=i}\left(\frac{\dot Q_i\dot Q_k}{Q_i(Q_i-Q_k)}+\frac{\hat v_k\hat w_i\hat Z_{k,i}}{Q_i}
+\frac{\dot Q_i\dot Q_k}{Q_k(Q_k-Q_i)}+\frac{\hat 
v_i\hat w_k\hat Z_{ik}}{Q_k}\right)\\
&=&\sum_{k\not=i}\frac{\dot Q_i\dot Q_k(Q_i+Q_k)-(Q_i-Q_k)(Q_i \hat v_i\hat w_k \hat Z_{ik}-Q_k \hat v_k\hat w_i\hat Z_{ki})}{Q_iQ_k(Q_i-Q_k)}
\label{eqn:genmotion}
\end{eqnarray}

\subsection{A Special Case}

We can further simplify \eqref{eqn:genmotion} assuming that $\hat w$
has no zero component.  In that case, we can utilize additional
freedom of conjugation by a diagonal matrix to leave $Q$ unchanged but
modify $U$.

In particular, if $\hat w$ is a vector \relax{with no zero component},
then we can put it in a form where
$w=(1,1,1,\ldots,1)$ by multiplying $U$ by the diagonal matrix with
$w_i$'s along its diagonal.  Now, in this ``gauge'', we know that
$\hat w_i=1$ and so by \eqref{eqn:dotqs} we know that $\hat v_i=\dot
Q_i$.  This then gives us that
$$
\ddot q_i=
\sum_{k\not=i}\frac{\dot Q_i\dot Q_k(Q_i+Q_k)-(Q_i-Q_k)(Q_i \dot Q_i \hat Z_{ik}-Q_k \dot Q_k \hat Z_{ki})}{Q_iQ_k(Q_i-Q_k)}.
$$

Ideally, we would like to be able to completely eliminate $\hat
Z_{ki}$ from this equation and have an ``intrinsic'' equation for the
eigenvalues.  It seems that this can only be done when certain additional
simplifying assumptions are made.

Suppose that we are in the case that
\begin{equation}
-\lambda \hat Y+\hat Z=\gamma I\Rightarrow 
\hat Z_{ij}=\lambda \hat Y_{ij} \ (i\not =j).\label{eqn:linear2}
\end{equation}
Combining equations \eqref{eqn:linear1} and \eqref{eqn:linear2} we find that
$$
\hat Z_{ij}=\frac{\lambda \hat v_j\hat w_i}{\lambda Q_i+Q_j}\qquad(i\not=j).
$$
Substituting this into \eqref{eqn:genmotion} and again using
\eqref{eqn:dotqs} one finds the intrinsic equations of motion
\begin{equation}
\ddot q_i=(\lambda-1)^2 \dot Q_i \sum_{k\not=i}\frac{\dot Q_k(Q_i+Q_k)}{(Q_i-Q_k)(\lambda Q_i-Q_k)(\lambda Q_k-Q_i)}
.\label{eqn:motion-lamgam}
\end{equation}

Note that the equations are independent of $\gamma$.  
In the case $\lambda=-1$ the dynamics of \eqref{eqn:motion-lamgam}
\textit{is} the Ruijsenaars-Schneider model \bibref[Braden].

\section{Comments and Conclusions}

It is interesting to note that
restrictions on $\k(X,Y,Z)$ for triples
of square matrices have arisen before in the context of integrable
systems.  For example, though the notations are different, the key
operator identity in \bibref[sakh] is such a restriction.  Perhaps
there is a deep connection between the results of that work and this
one, though the relationship is not immediately apparent to us.
A more similar result was obtained in \bibref[NC] where the
condition $\textup{rank}(XZ-qZX)=1$ for invertible $X$ and $Z$ and
scalar $q$ was used to construct solutions to the $q$-difference KP
hierarchy.  
  It is reasonable to suppose that their result could now
also be obtained as a discretization of the results in the present
work in the special case $Y=qZ$.  (Another matrix approach to $q$-KP
\bibref[Iliev] made use of the condition $\textup{rank}(XY-qYX+I)=1$.)

The suggestive appearance of these spaces of matrices in such
different contexts within the study of integrable systems might
indicate that we should look more carefully at the manifolds $\M_n^k$.
For instance, we have implicitly constructed a map from
$\M_n^1$ to the infinite dimensional grassmannian $Gr_1$ \bibref[SW],
and $\M_n^k$ naturally has the structure of an algebraic variety, but
so far we have little understanding of the geometry.

In \bibref[W2], Wilson constructs an adelic grassmannian and a Hilbert
scheme from the set of matrices satisfying
$\textup{rank}([X,Z]+I)=1$.  Moreover, the natural symmetry of this
set which is manifested as the involution
$(X,Z)\mapsto(Z^{\top},X^{\top})$ has significance both for the KP
hierarchy (bispectrality) and the Calogero-Moser particle system
(self-duality).  So, it is reasonable to wonder how the obvious
symmetries of $\M_n^k$ are reflected in the soliton solutions to the
KP hierarchy.  We have already noted that multiplying $X$ by a
function of $Y$ on the left and a function of $Z$ on the right
corresponds to the KP flows.  Note also that if $\k(X,Y,Z)=1$ and $X$
is invertible then $\k(X^{-1},Z,Y)=1$ as well and that this triple
corresponds to the same KP solution.  (In particular, these two points
in $\M_n^1$ get mapped to the same point in $Gr^{rat}$.)  Similarly,
if $Y$ is invertible then $\k(Y,X,XZY^{-1})=1$, but it is not
immediately apparent what symmetry of KP is analogous.

One alternative characterization of $Gr^{rat}$ is as the grassmannian
of finite dimensional subspaaces of finitely supported distributions
\bibref[cmbis].  Specifically, to identify a point $W\in Gr^{rat}$ it
is sufficient to identify the finitely supported distributions in $z$
which annihilate the normalized Baker-Akhiezer function. We showed in
Section~\ref{sec:solitons} that in the case of non-degenerate
solitons, the eigenvalues of $Y$ and $Z$ determine the support of the
distributions and $X$ determines the coefficients.
We conjecture that this situation holds in general: 

\begin{conjecture}{The support of
the distributions annihilating $z^n\psi_M$ for $M=(X,Y,Z)\in\M_n^1$ is
the set of eigenvalues of the matrices $Y$ and $Z$ with the highest
derivative taken at a particular eigenvalue being bounded by the size
of the corresponding Jordan block.}
\end{conjecture}

\bigskip\noindent\textbf{Acknowledgement:}  We are grateful to Annalisa Calini, Harry
Braden, John Harnad, and Tom Ivey for helpful discussions.

\end{document}